\documentclass{aastex61}

\usepackage[encapsulated]{CJK}
\usepackage{ucs}
\usepackage[utf8x]{inputenc}
\newcommand{\cntext}[1]{\begin{CJK}{UTF8}{gbsn}#1\end{CJK}}

\newcommand\aastex{AAS\TeX}

\received{September 28, 2017}
\revised{xx xx, 2017}
\accepted{xx xx, 2017}

\shorttitle{\aastex\ manuscript article}
\shortauthors{Jiang et al.}

\begin{document}

\title{VLBI imaging of M81* at $\lambda$ = 3.4\,mm with source-frequency phase-referencing}

\correspondingauthor{Wu Jiang \& Zhiqiang Shen}
\email{jiangwu@shao.ac.cn, zshen@shao.ac.cn}

\author[0000-0002-0786-7307]{Wu Jiang (\cntext{江悟})}
\affiliation{Shanghai Astronomical Observatory, 
 Chinese Academy of Sciences, 
 Shanghai 200030, China}
\affiliation{Key Laboratory of Radio Astronomy,  
 Chinese Academy of Sciences, Nanjing 210008, China}
\affiliation{University of Chinese Academy of Sciences,  
  Beijing 100049, China}

\author{Zhiqiang Shen (\cntext{沈志强})}
\affiliation{Shanghai Astronomical Observatory, 
 Chinese Academy of Sciences, 
 Shanghai 200030, China}
\affiliation{Key Laboratory of Radio Astronomy,  
 Chinese Academy of Sciences, Nanjing 210008, China}

\author{Dongrong Jiang (\cntext{蒋栋荣})}
\affiliation{Shanghai Astronomical Observatory, 
 Chinese Academy of Sciences, 
 Shanghai 200030, China}
\affiliation{Key Laboratory of Radio Astronomy, 
 Chinese Academy of Sciences, Nanjing 210008, China}

\author{Ivan Mart\'i-Vidal}
\affiliation{Department of Space, Earth and Environment, 
Chalmers University of Technology, 
Onsala Space Observatory, SE-43992, Sweden}

\author{Noriyuki Kawaguchi}
\affiliation{Shanghai Astronomical Observatory, 
 Chinese Academy of Sciences, 
 Shanghai 200030, China}



\begin{abstract}

We report on the first VLBI image of the M81 nucleus (M81*) at a wavelength of 3.4\,mm, obtained with the source-frequency phase-referencing (SFPR) technique. Thanks to the SFPR calibration, the coherent integration time could be eventually increased by more than an order of magnitude, which enabled the detection of fringes at the level of 45\,mJy\,beam$^{-1}$ with a dynamic range higher than 130:1. This paves the way toward future mm/sub-mm VLBI observations of weaker sources. From the analysis of the M81* visibilities, a core size of $\sim50$ $\mu$as at 3.4\,mm was estimated. This follows the power-law relationship with wavelength, $\propto \lambda^{0.88}$, reported previously at lower frequencies. These results constrain the core size (at 3.4\,mm) to a minimum of $\thicksim$80 Schwarzschild radii of M81*.

\end{abstract}




\section{Introduction} \label{sec:intro}
The active nucleus of galaxy M81 (hereafter M81*), at a distance of $3.96\pm0.29$\,Mpc \citep{2007ApJ...668...924B}, is one of the nearest low luminosity active galactic nuclei, LLAGN \citep{2002A&A...392..53N}. With a black hole mass of $\thicksim7\times10^7M_\odot$ determined by spectroscopy of Hubble Space Telescope \citep{2003AJ...125...1226D}, an angular resolution of 1\,$\mu$as corresponds to a linear size of 3 Schwarzschild radii ($R_s$) on M81*. Discrete knot ejections from its jet were discovered in a multi-wavelength campaign \citep{2016NatPh...12..772K}. M81* provides an ideal target for studying the physics of weakly-accreting black holes and the jet formation in LLAGN. It also serves as an apparent transitional object between Sgr A* \citep{2005Natur...438..62S,{2006A&A...451...845B}} and high luminosity AGN. VLBI observations of M81* at wavelengths from 18 cm to 7 mm have been carried out for more than a decade. As seen with VLBI, M81* was identified with a compact core and a one-sided jet to the northeast \citep{2000ApJ...532...895B, 2004ApJ...615...173B}. The apparent source size is about 0.5\,mas ($\thicksim0.01$\,pc) at a wavelength of 3.5\,cm, and follows a power-law relationship with wavelength ($\lambda^{0.88}$) between 7\,mm and 13\,cm \citep{2012A&A...537..A93R}. Besides, the position angle of the jet structure in M81* seems to be wavelength dependent, changing clockwise from $\thicksim75^{\circ}$ at 13\,cm to $\thicksim28^{\circ}$ at 7\,mm. A long-term global VLBI monitoring of M81* also showed strong evidence of jet precession towards the south \citep{2011A&A...533..A111M,2013EPJWJ...61...08002A}.

Although M81* is an interesting source to study the LLAGN, unfortunately it is not possible to directly observe with high-resolution 3\,mm VLBI. The coherence  time at this wavelength is of a few tens of seconds, at most, limiting the positive detection of fringes with the current VLBI sensitivity. The nearest strong calibrator B0954+658 is $3.5^{\circ}$ away from  M81*. Since the telescope slewing times needed for conventional phase referencing are relatively long, it is impossible to properly calibrate the atmospheric phase fluctuations on M81* at 3\,mm. However, it was firstly introduced by \citet{2005A&A...433..897M} that fast frequency-switching phase-referencing can be used to calibrate tropospheric phase fluctuations and eventually increase the integration time. Unlike the traditional VLBI phase referencing observation, the phase corrections for the target are referenced to the lower frequency phases of the same target source. The phase gains at the lower reference frequency, or simply reference frequency ($\nu_r$), are scaled by the frequency ratio ($\gamma$) and interpolated to the phase gains at the higher target frequency, or simply target frequency ($\nu_t$). The tropospheric phase errors and the antenna position errors are non-dispersive and can be removed by the phase gains. The ionospheric phase errors are dispersive but can be reduced by ionospheric model derived from GPS data at the same time period. Furthermore, the unmodeled ionospheric delay and the instrumental phase offsets between the two frequencies ($\nu_r$ and $\nu_t$) can be further eliminated by calibrator observations. Such an improved technique was named source-frequency phase-referencing (SFPR) \citep{2011AJ....141..114R}. After the SFPR calibration, the remaining phases just reflect the true high frequency visibility and the frequency-dependent shift in the positions of the cores (core shift) \citep{2008A&A...483..759K}. Following Eq. 7 in \citet{2011AJ....141..114R}, we have
\begin{equation}
\phi^{SFPR}_{tar} = \phi^{tru}_{tar, \nu_t} + 2\pi \boldmath {D_{\nu_t}} \boldmath \cdot (\boldmath \theta_{tar} - \boldmath \theta_{cal})+ {\Delta _{i,T^{\nu}_{swt}} + \Delta _{i,T_{swt}}},
\end{equation}
where, $\phi^{SFPR}_{tar}$ is the SFPR-calibrated (SFPR-ed) phase of the target, $\phi^{tru}_{tar, \nu_t}$ is the phase of the true high frequency visibility, $\boldmath D_{\nu_t}$ is the interferometer baseline in units of wavelength of $\nu_t$, $\boldmath{\theta_{tar}}$ and $\boldmath{\theta_{cal}}$ are the core shift between the two frequencies ($\nu_t$, $\nu_r$) of the target source and calibrator source respectively, $\Delta _{i,T^{\nu}_{swt}}$ is the interpolation errors from a frequency-switching cycle $T^{\nu}_{swt}$, and $\Delta _{i,T_{swt}}$ is the interpolation errors from a source-switching cycle $T_{swt}$. To present the two types of  the SFPR calibration including both the  phase gains of the reference frequency and the phases from the calibrator source, the left side of Eq. 1 is expanded and the SFPR calibration can be expressed as:
\begin{equation}
\phi^{obs}_{tar, \nu_t} - \gamma \hat \phi^{int}_{tar, \nu_r} -(\phi^{obs}_{cal, \nu_t} - \gamma \hat \phi^{int}_{cal, \nu_r}) = \phi^{tru}_{tar, \nu_t} + 2\pi \boldmath D_{\nu_t} \boldmath \cdot (\boldmath \theta_{tar} - \boldmath \theta_{cal})+\Delta \phi_n,
\end{equation}
where, $\phi^{obs}_{tar, \nu_t}$ and $\phi^{obs}_{cal, \nu_t}$ are the visibility phases of the target and the calibrtor at the high frequency respectively, $ \hat \phi^{int}_{tar, \nu_r}$ and $\hat \phi^{int}_{cal, \nu_r}$ are the interpolated phases of the target and the calibrtor at the reference frequency respectively, whilst the calibrator is assumed point like or calibrated by its clean components, and $\Delta \phi_n$ includes the two types of interpolation errors. 
In this Letter, we present the first VLBI image of M81* at a wavelength of 3.4\,mm with SFPR.
\section{Observations} \label{sec:obs}
We observed M81* on February 27, 2016 with 8 VLBA stations at the wavelengths of 7\,mm and 3.4\,mm. To avoid phase wraps in SFPR, the target frequency should be an integer multiple of the reference frequency. In the case of our VLBA observations, we chose the 43860 and 87720\,MHz pair with an integer multiplier of 2 as the reference frequency and the target frequency, respectively. The overall observing time was 8 h. A bandwidth of 256\,MHz in both left-hand circular polarization (LCP) and right-hand circular polarization (RCP) was recorded with a 2-bit sampling at a data rate of 2048\,Mbps. The 256\,MHz bandwidth was splitted into 8 IFs with 32 MHz bandwidth each. Each IF was further subdivided into 64 spectral channels and integrated by two seconds in the correlator. Dynamic scheduling allowed us to observe under good weather condition at most stations. All antennas performed well, except for Owens Valley (OV) and North Liberty (NL), where OV got a high system temperature for the 3.4\,mm receiver in RCP, and NL site was windy during the observation and thus its performance at 3.4\,mm was seriously affected. Data were correlated at Socorro with the DiFX correlator \citep{2011PASP...123...275D}.

The BL Lac object B0954+658 ($3.5^{\circ}$ apart from M81*) was chosen as the phase and amplitude calibrator, QSO 1044+719 ($5.2^{\circ}$ apart) was used as the polarization calibrator, the strong calibrator OJ287 served as the fringe finder and D-term calibrator. Since the observation was scheduled in SFPR mode, a special strategy was considered on the scan nodding. B0954+658 was interleaved every 10 or 14 loops of M81*, depending on the elevation angle. Each loop of M81* included a 30 second scan at 7\,mm and a 60 second scan at 3.4\,mm. A relatively longer cycle period was selected because of a high elevation angle (mostly $40^{\circ}-60^{\circ}$) on M81* at 7 of the 8 VLBA stations and a clear weather requested for the 3 mm observation.  
\section{Data reduction} \label{sec:red}
The data reduction followed standard procedure to the amplitude and phase calibrations in AIPS. The amplitudes were calibrated using system temperatures and gain curves provided by stations, and amplitude corrections for errors of 2-bit-sampling were performed using autocorrelation data. The bandpass correction was derived at each frequency from the fringe finder scan. Phase corrections for parallactic angles were applied. The ionospheric correction was conducted by the AIPS task TECOR with ionospheric total electron content maps from the Jet Propulsion Laboratory. Instead of using the pulse calibration system at stations, the fringe-fitting results of the fringe finder scans at each frequency were used to correct the instrumental delays and phase offsets among different IFs.

The main process of the data reduction, the search for the group delay and phase rate, had to be performed in several steps. The first step was to obtain an image of M81* at 7\,mm. A fringe-fitting was performed by combining all the IFs to increase signal to noise ratio (SNR). The SNR threshold of 4 was selected to obtain more solutions especially from Mauna Kea (MK) related baselines, whose SNRs were generally lower due to their longer baseline lengths than those of the continental baselines. About 90\% good solutions were obtained on M81*, the results were applied and exported to Difmap for final imaging. Once we had the deconvolved image of M81* at 7\,mm, the clean components resulting from Difmap were imported back to AIPS as an input for the new round of fringe-fitting. In this step, the delay search window was narrowed to 1 nanosecond (ns) and the rate was not fitted. An SNR of 5 was used to extract the phase and delay solutions in this run of FRING. As a result, about 95\% good solutions were obtained, guaranteeing the completeness of the phase connection in later processing. The 7\,mm solutions table was exported to ASCII format with the AIPS task TBOUT, so that the scaling of phase gains with frequency ratio could be performed outside the AIPS environment. The 7\,mm phase gains of M81* at the 8 VLBA stations, which are to be applied to the 3.4 mm data after being multiplied by 2, are shown in Figure 1(a). The phases change smoothly at all the stations except some periods at OV station. We kept the LCP and RCP gain solutions independent, during the whole SFPR process, as an additional check for self-consistency (since, once the instrumental RCP/LCP offset is calibrated using the fringe finder, the phase and rate gains should always be the same for both polarizations). We also fringe-fitted the 3.4\,mm data of M81*, using a low SNR threshold of 3 and a solution interval of 1 minute. This was done to compare the quality of SFPR with that of an ordinary calibration. In the case of fringe-fitting on  the 3.4\,mm data, the success rate (and quality) of the solutions was very limited (see Fig. 1b), even if polarizations were combined in the fringe search. Hence, integration times longer than 1 minute (only achievable with SFPR) are needed for a clear detection of M81* fringes at 3.4\,mm. A Python program was developed to scale the 7\,mm phases by a factor of two while keeping the delay unchanged. The new solutions table was imported back into AIPS with TBIN and used as a solution table at the target frequency. Both the phase interpolation and connection were conducted with AIPS task CLCAL. A re-fringe fitting procedure was performed on the B0954+658 data at 3.4\,mm, pre-calibrated by the 7\,mm phase gains. These solutions were then applied to M81*, to further refine the calibration of unmodeled ionospheric variations and instrumental offsets between 7\,mm and 3.4\,mm (see Eq. 2). Hence, the SFPR calibrations were finished and the SFPR-ed dataset of M81* was obtained.

Before making a final image from the SFPR-ed 3.4\,mm visibilities of M81*, the  phase gains (see Fig. 1c) and residual delays were obtained by a second run of fringe-fitting, with a solution interval as along as the time span of 10 or 14 loops of M81* scans. The re-fringe fitting was made with an SNR threshold of 4. We examined the fringe SNR of M81* at 3.4\,mm from FRING during the re-fringe fitting procedure. Figure 1(d) shows the mean fringe SNRs of Los Alamos (LA) related baselines over different integration times. As expected, an improvement in the SNR with increasing integration time can be clearly seen. The maximum integration time at 3.4\,mm with SFPR could be more than 20 minutes in our observations. The SNR of about 3.5 is also reasonable on the short baselines such as LA-Fort Davis (FD), LA-Kitt Peak (KP) and LA-Pie Town (PT) with an integration time of about 460 seconds. The theoretical sensitivity of these baselines at 3.4\,mm is around 30\,mJy in a 30 second integration time with 2048\,Mbps recording rate. The result is also consistent with the flux density of M81* at 3.4\,mm in our image. The longer baselines, LA-Brewster (BR) and LA-MK, showed a lower SNR due to M81* of being partially resolved. 
\section{Results and discussions} \label{sec:results}
\subsection{VLBI image of M81* at 3.4\,mm}\label{subsec:image}
After re-fringe fitting with a long integration time enabled by the application of SFPR  on M81* at 3.4\,mm, about 65\% good solutions were obtained. The failed solutions were mainly related to the two stations, NL and OV. In the final imaging processing, no NL station data was included and just 25\% of OV station data was kept. The visibility data of B0954+658 at 3.4\,mm was calibrated and imaged separately. A preliminary amplitude correction on M81* was made by using the amplitude calibration solutions of B0954+658 every 30 minutes. It helped to compensate the pointing loss and instrumental variations. The peak intensity was 26.6\,mJy\,beam$^{-1}$ and the root mean square (RMS) noise was 1.47\,mJy\,beam$^{-1}$ in the dirty image, yielding a dynamic range of 18:1. Since the global fringe-fitting was in a time scale of tens of minutes, minor phase variations should be calibrated by closure quantities. The visibility data were averaged in bins of 58 seconds. Several iterations of phase self-calibration were executed, prior to an amplitude self-calibration with a solution interval of 30 minutes. Both the phase and amplitude self-calibration contributed to the flux recovery. The final deconvolved image had a peak intensity of $\thicksim$45\,mJy\,beam$^{-1}$ and an RMS noise of 0.34\,mJy\,beam$^{-1}$. Its dynamic range was improved to more than 130:1 (Fig. 2). The theoretical RMS at 3.4\,mm was about 0.31\,mJy\,beam$^{-1}$. Given the average time of 58 seconds in the phase self-calibration and 2 seconds in the correlator, the amount of spurious source flux density induced by phase self-calibration in our observations can be calculated using Eq.~7 in \citet{2008A&A...480..289M}; its level is as low as 10\% of the RMS of the visibilty amplitude, which ensures a robust self-calibration. The CLEAN beam of our final image is 0.17\,mas$\times$0.101\,mas.

We model-fitted the calibrated visibility data at both 7\,mm and 3.4\,mm with Gaussian functions. A two-component model was used to fit the core and the jet emission for the 7\,mm data. The 3.4\,mm data was fitted by two different fitting procedures, a single elliptical Gaussian and two circular Gaussians. The results are listed in Table 1. We took the standard deviation for the uncertainty estimation. From the fit of two Gaussian components, the jet emission at 0.33\, mas from the core with a position angle of $57^{\circ}$ was detected at 7\,mm, while previous 7\,mm VLBA observations in 2002 located the jet emission at the same distance with a position angle of $50^{\circ}$. A $\thicksim0.5^{\circ}$ per year of change in the position angle is consistent with the long-term drift derived in \citet{2011A&A...533..A111M}. The observed evolution of the jet at 7\,mm needs to be confirmed by future observations because of the lacking of 7\,mm VLBI observations between these two epochs ($\thicksim$ 14 years). However, no obvious jet emission at 3.4\,mm could be seen, while the core accounted for more than 90\% of the total emission.

Figure 3 shows the core size of M81* as a function of wavelength by combining previously published data up to 7\,mm \citep{1996ApJ...532...895B,2011A&A...533..A111M,2012A&A...537..A93R} with our new data at 7\,mm and 3.4\,mm. The core sizes of the previous results were averaged within each frequency. A direct conclusion of Figure 3 is that the model of core-size dependence with wavelength $\lambda^{0.88}$ \citep{2012A&A...537..A93R} is still valid at 3.4\,mm. A more thorough fitting to all the measured sizes from 3.4\,mm to 17.6\,cm  leads to a power-law scheme of $\lambda^{0.89\pm 0.03}$  and $\lambda^{0.87\pm 0.06}$ for the major axis and minor axis, respectively. Such a dependence of core size on the observing wavelength is consistent with the inhomogeneous relativistic jet model \citep{1998ApJ...494..139J}. It can be seen that the angular size of M81* at 3.4\,mm is in a range of 61 to 26\,$\mu$as. This constrains the core emission region at 3.4\,mm to a minimum size of  $\thicksim$80 $R_s$.
\\
\\
\subsection{Core shift and spectral index map of M81*} \label{subsec:spec}
The core shift of M81* at 3.4\,mm was derived from the SFPR-ed image of M81* (Fig. 4: left panel), which was obtained before re-fringe fitting on M81*. The image was obtained through a global clean using the IMAGR task in AIPS without any self-calibration. Its convergence to a single clean component indicated that the SFPR technique was successfully applied. The JMFIT task in AIPS was used to extract the astrometric offset of peak position with respect to the phase center. The offset was $-35\pm2$ $\mu$as in right ascension and $-32\pm3$ $\mu$as in declination, respectively. We noticed that a core shift in B0954+658 will be transferred to M81* as indicated in Eq. 1 and Eq. 2. Using the estimated core shift of B0954+658 from previous publication \citep{2009MNRAS...400..26O}, it was possible to separate the core shift of M81* from the SFPR-ed image. Here the core shift measurement ($0.18\pm0.02$\,mas, $127.3\pm0.6^{\circ}$) of B0954+658 from 4.6-8.9\,GHz was adopted and equipartition in the radio core from 4.6-88\,GHz was assumed. As the jet itself is assumed to have no transverse velocity gradient and the frequency dependence of observed position of the VLBI core is along the jet direction \citep{1998A&A...330..79L}, the position angle of the shift of $\thicksim127^{\circ}$ was consistent with the 43\,GHz inner jet direction of  $\thicksim-59^{\circ}$ in \citet{2009MNRAS...400..26O}. In our observations, the position angle of the inner jet at 7\,mm was changed to $-$23$^{\circ}$ and the core shift direction of B0954+658 was then assumed to be $157^{\circ}$ for consistency. The derived offset of M81* has a magnitude of 57$\pm14$ $\mu$as with a direction of $ -151\pm5^{\circ}$, consistent with the jet direction and former prediction \citep{2004ApJ...615...173B}. The core shift uncertainty was conservatively estimated to be given by the full-width at half maximum of the beam divided by the peak component's SNR. If we adopt the position angle of $127.3^{\circ}$, the core shift of M81* would be 48 $\mu$as at a direction of 156$^{\circ}$ and is still within the error range of our results. The spectral index map in Figure 4 (right) was made by convolving 3.4\,mm data with the 7\,mm beam. The AIPS task COMB was used to calculate the spectral index across the image. The pixels, whose values were below three times the residual noise level, were clipped to ensure a real detection. Since part of the emission at 3.4\,mm was resolved out, the spectral index map shall be interpreted as a lower limit (i.e., the true spectrum might be flatter). The spectral index map shows that our estimate of the core shift is reasonable, given the relatively flat spectrum at the core region, as expected \citep{2008ApJ...681..905M}, together with a spectral gradient in the direction of the jet. Since the intensity of the jet extension was very weak at 3.4\,mm, the spectral index at that region becomes very steep (the pixels in that region are indeed masked, due to the noise clipping at 3.4\,mm). However, this spectral index map is still preliminary. Our results should be refined by accurately aligning the optically thin emission or by a more precise core shift measurement in future observations.
\section{SUMMARY} \label{sec:sum}
We have successfully applied the SFPR technique on VLBI observations of the LLAGN in M81 (M81*). After calibration of the 3.4\,mm visibilities from the phase gains at 7\,mm, the coherent integration time of the 3.4\,mm visibilities was increased by more than an order of magnitude, hence allowing to obtain clear detections of M81* at 3.4\,mm. The nucleus of M81, whose peak flux density was around 45\,mJy, was imaged with a dynamic range better than 130:1. This is the first VLBI image of M81* at 3.4\,mm, which sets a constraint of the core size to a minimum of  $\thicksim$80 $R_s$. M81* had a relatively low flux density and might stay in the quiet phase during our observations. The jet emission at 3.4\,mm was resolved out to three times the noise level and could not be firmly detected. The detection of jet and knot components should be possible in future observations during the active variable phase of M81*. The SFPR technique has a great advantage for correcting tropospheric phase fluctuations and increasing the integration time at mm-VLBI observations. It is beneficial to the detection and imaging of weak sources at mm/sub-mm wavelength. The data reduction scheme reported here will also be referential to a simultaneous multi-frequency VLBI receiving system \citep{2013PASP..927..539H}.

We would like to thank the anonymous referee for his/her very critical and constructive suggestions. This work was supported in part by the Major Program of the National Natural Science Foundation of China (Grant No. 11590780, 11590784) and Key Research Program of Frontier Sciences, CAS (Grant No. QYZDJ-SSW-SLH057). VLBA is operated by the National Radio Astronomy Observatory, which is a facility of the National Science Foundation operated under cooperative agreement by Associated Universities, Inc.

\begin{figure*}
\gridline{\fig{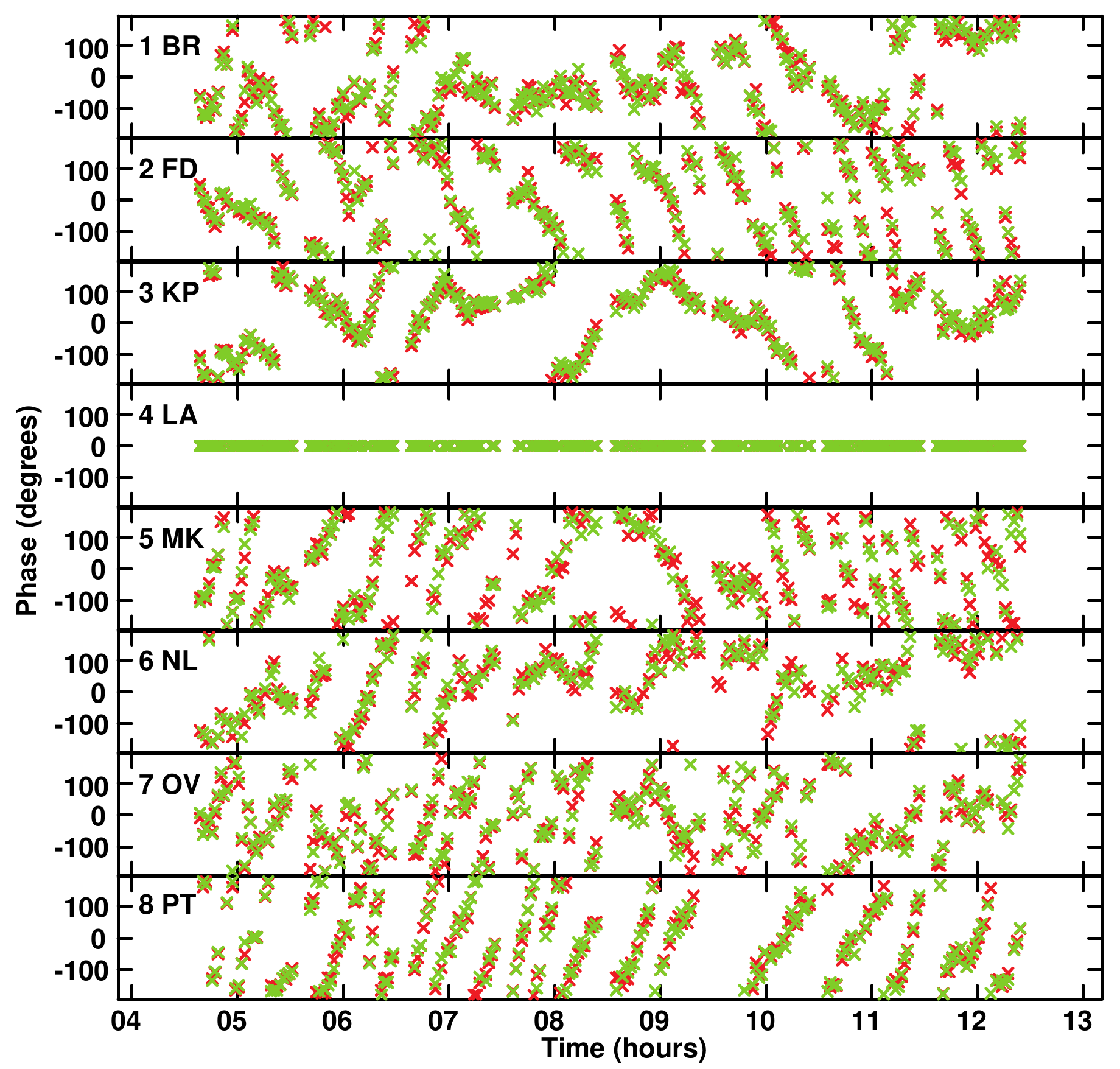}{0.5\textwidth}{(a)}
          \fig{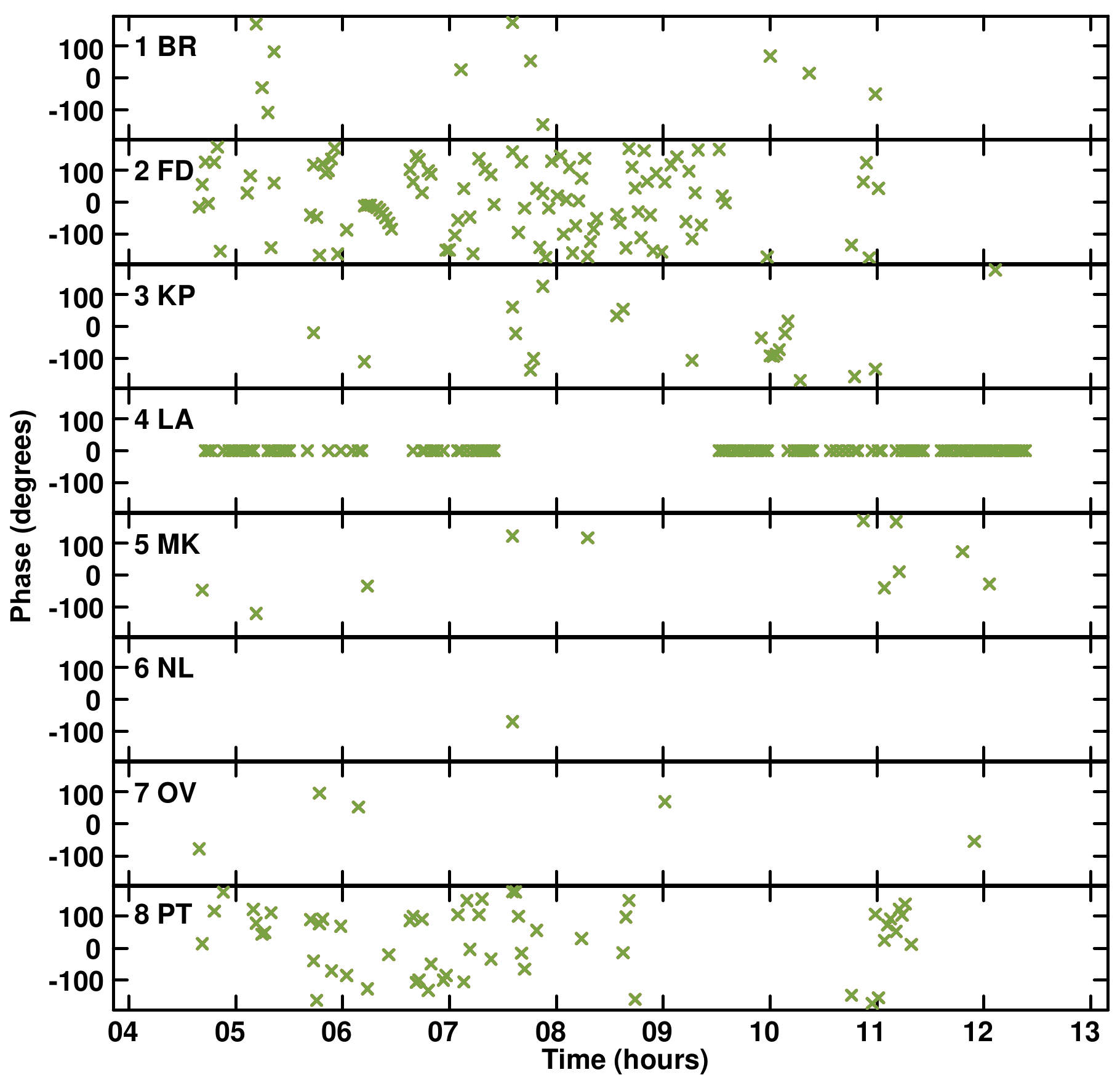}{0.5\textwidth}{(b)}
          }
\gridline{\fig{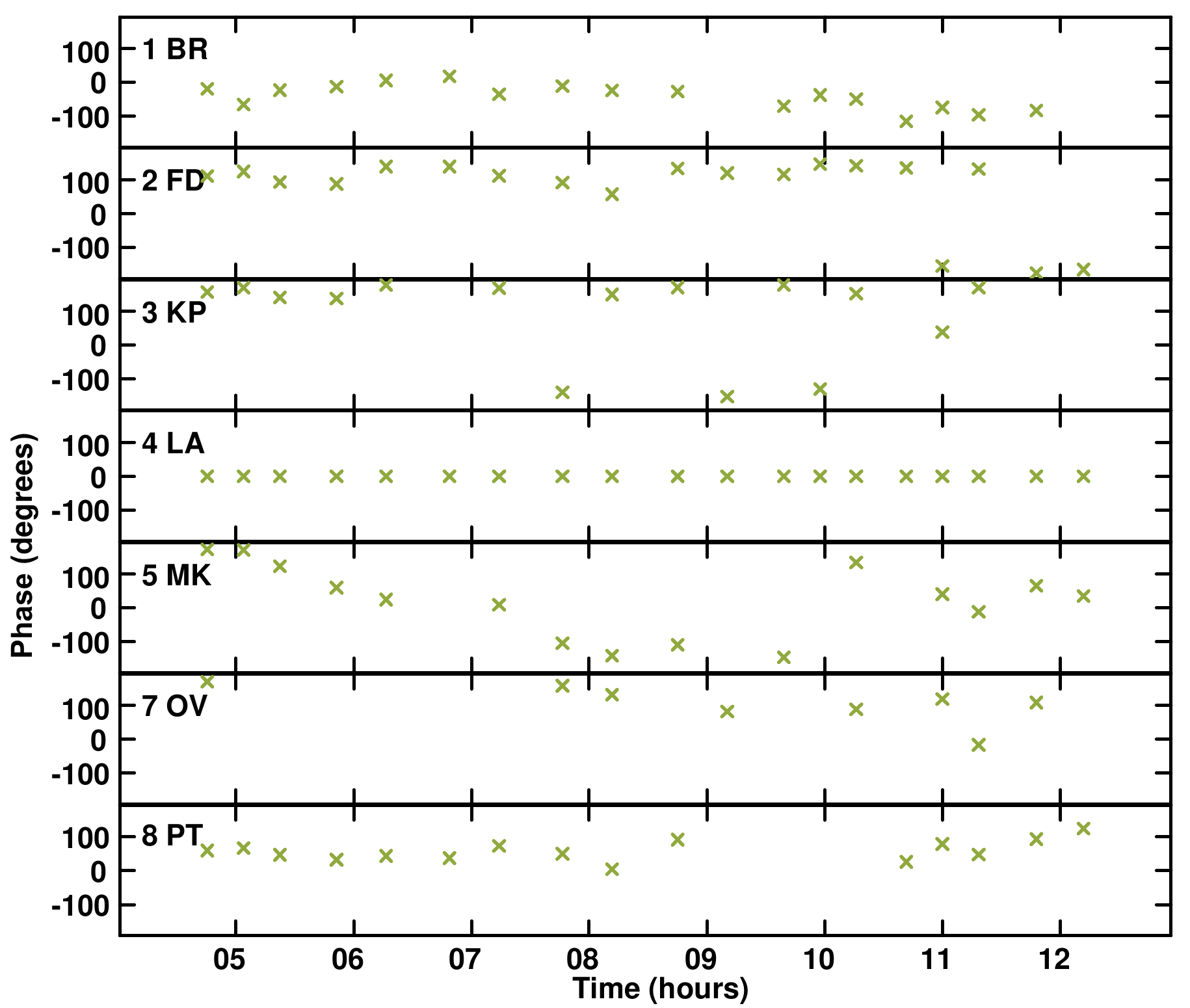}{0.5\textwidth}{(c)}
          \fig{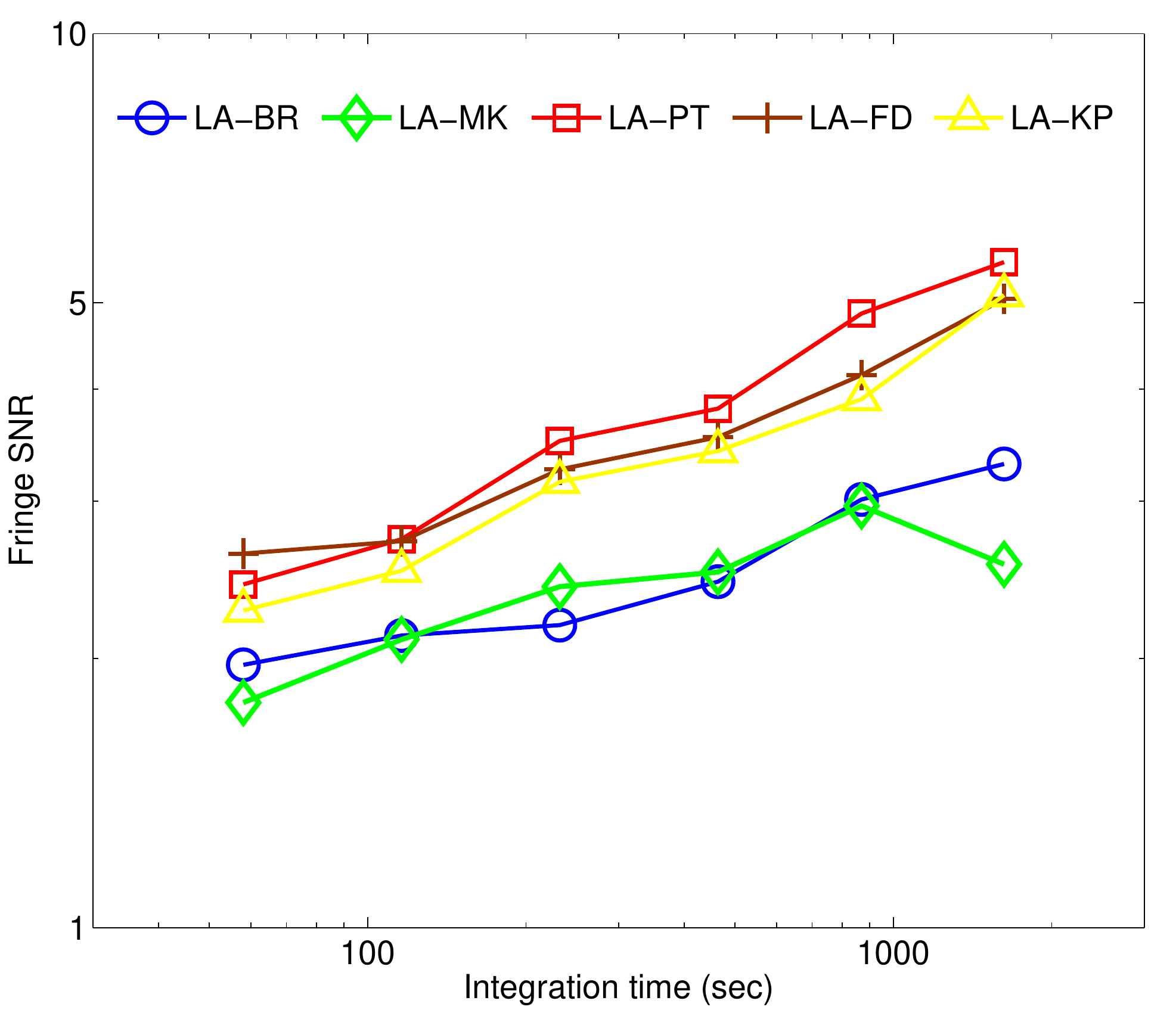}{0.5\textwidth}{(d)}
          }
\caption{(a) The phase gains of M81* at the 8 VLBA stations from fringe-fitting directly on the 7\,mm data. Green symbol corresponds to the phase in LCP; red symbol to the phase in RCP. (b)  The phase gains of M81* at the 8 VLBA stations from fringe-fitting directly on the 3.4\,mm data. (c)  The 3.4\,mm phase gains of M81* at the 7 VLBA stations (NL excluded due to no valid solution) from fringe-fitting on the SFPR-ed data. LA is the reference station in all the gains shown in this figure.  (d) The fringe SNR of M81* at 3.4\,mm with respect to integration time after phase corrections with SFPR. 
 \label{fig:fig1}}
\end{figure*}

\begin{figure}[h!]
\centering
\includegraphics[width=6in]{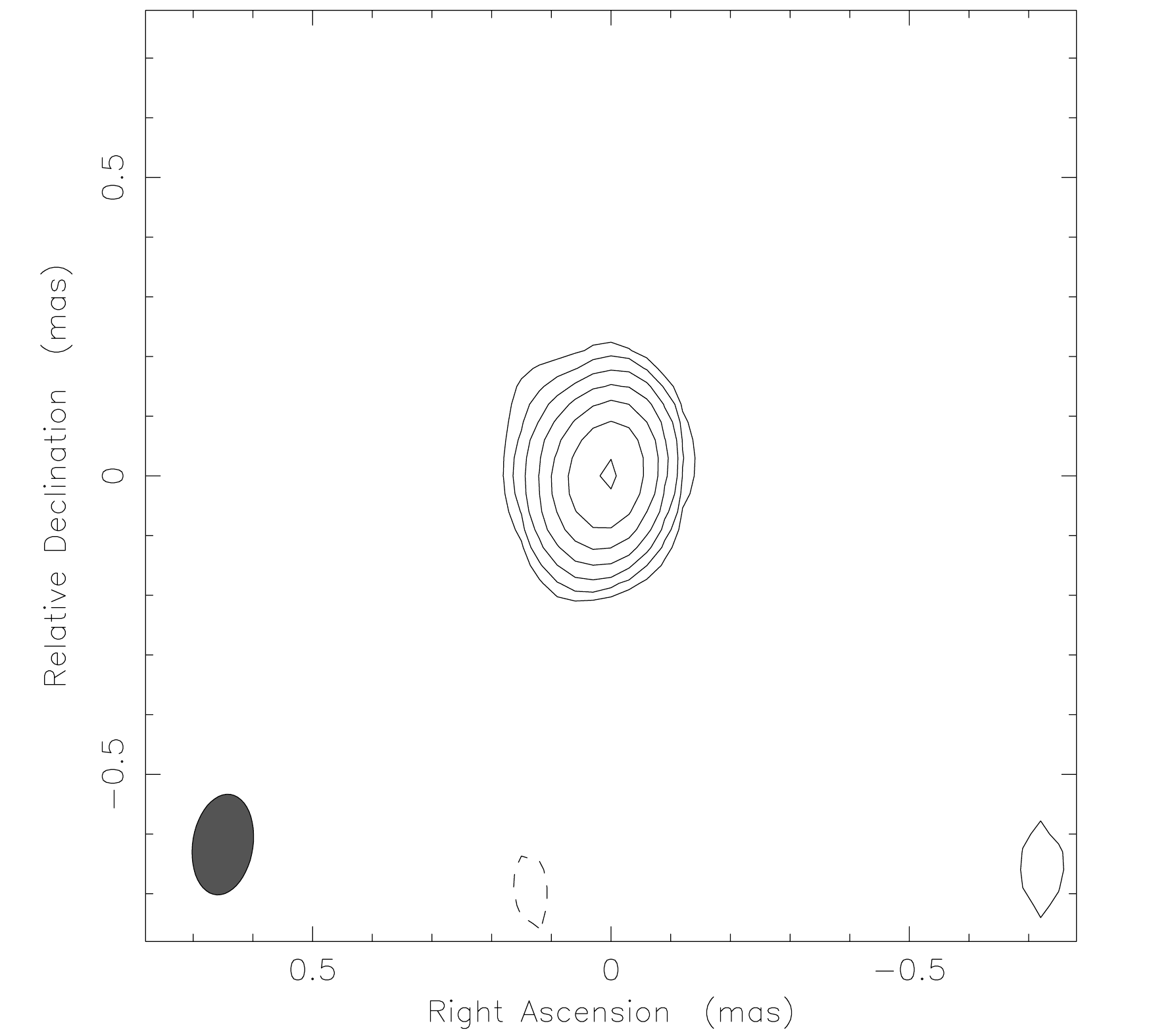}
\caption{VLBI image of M81* on 2016 Feburary 27 at 3.4\,mm. The peak intensity is 45.3 mJy\,beam$^{-1}$. The contour levels are 0.67$\times$($-$1, 1, 2, 4, 8, 16, 32, 64) mJy\,beam$^{-1}$. The lowest level is two times RMS noise of the image. The interferometric beam, shown at lower left, has a size of 0.17\,mas$\times$0.101 mas and a position angle $-$8.7$^{\circ}$. \label{fig:fig2}}
\end{figure}

\begin{deluxetable}{cccccccc}
\tablecaption{Gaussian model fit results \label{tab:mathmode}}
\tablecolumns{10}
\tablenum{1}
\tablewidth{0pt}
\tablehead{
\colhead{$\lambda$} & \colhead{Model} & 
\colhead{S} & \colhead{r} & \colhead{$\phi$} & \colhead{a} & 
\colhead{b/a} & \colhead{P.A.}  \\
\colhead{(mm)} & \colhead{} & \colhead{(mJy)} & \colhead{(mas)} & 
\colhead{(deg)}  & \colhead{(mas)} & \colhead{} & \colhead{(deg)}
}
\colnumbers
\startdata
{7.0}  & $1^{st}$ ellip. (core) & $98.9\pm0.5$ & ...             & ...        & $0.141\pm0.002$ &$ 0.28\pm0.07$ & $62\pm1.2$ \\
{   }  & $2^{nd}$ circ. (jet) & $10.8\pm0.5$  & $0.33\pm0.01$ &  $57\pm1.5$ & $0.199\pm0.015$ & 1            & ...         \\
\hline
{   }  & Single ellip.    & $53.6\pm0.8$  & ...             & ...         & $0.061\pm0.003$ & $0.42\pm0.20$&$89\pm7.0$ \\
{3.4}& $1^{st}$ circ. (core) & $50.2\pm2.0$  & ...             & ...         & $0.047\pm0.006$ & 1            & ...        \\
{   }  & $2^{nd}$ circ. (jet) & $4.0\pm2.0$   & $0.08\pm0.02$ &$115\pm20$ & $0.032\pm0.030$& 1            & ...         \\       
\enddata
\tablecomments{Col.(1) Observing wavelength in millimeter;  Col.(2) Gaussian functions of the model; Col.(3) Flux density in mJy; Col.(4) Distance from the phase center in mas; Col.(5) Position angle with respect to the phase center; Col.(6) Major axis of the fitted component in mas; Col.(7) Axis ratio of the fitted component; Col.(8) Position angle of the component's major axis.}
\end{deluxetable}

\begin{figure}[h!]
\centering
\includegraphics[width=6in]{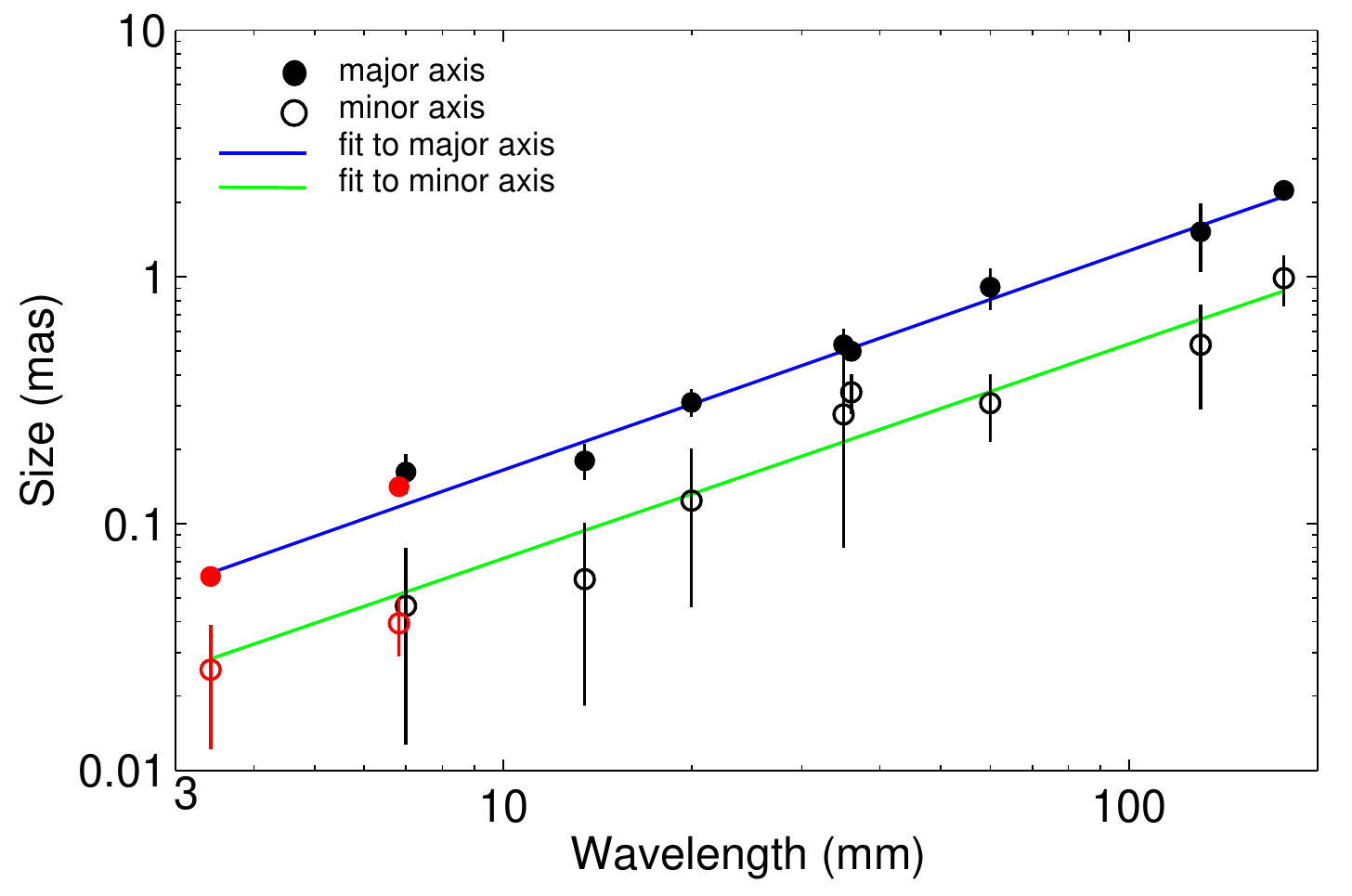}
\caption{Core size of M81* as a function of wavelength. Blue line corresponds to the major Gaussian axis fitted to the core; green line to the minor Gaussian axis. Black circles are the previous measurements, red circles are our new results. \label{fig:fig3}}
\end{figure}

\begin{figure}
\plottwo{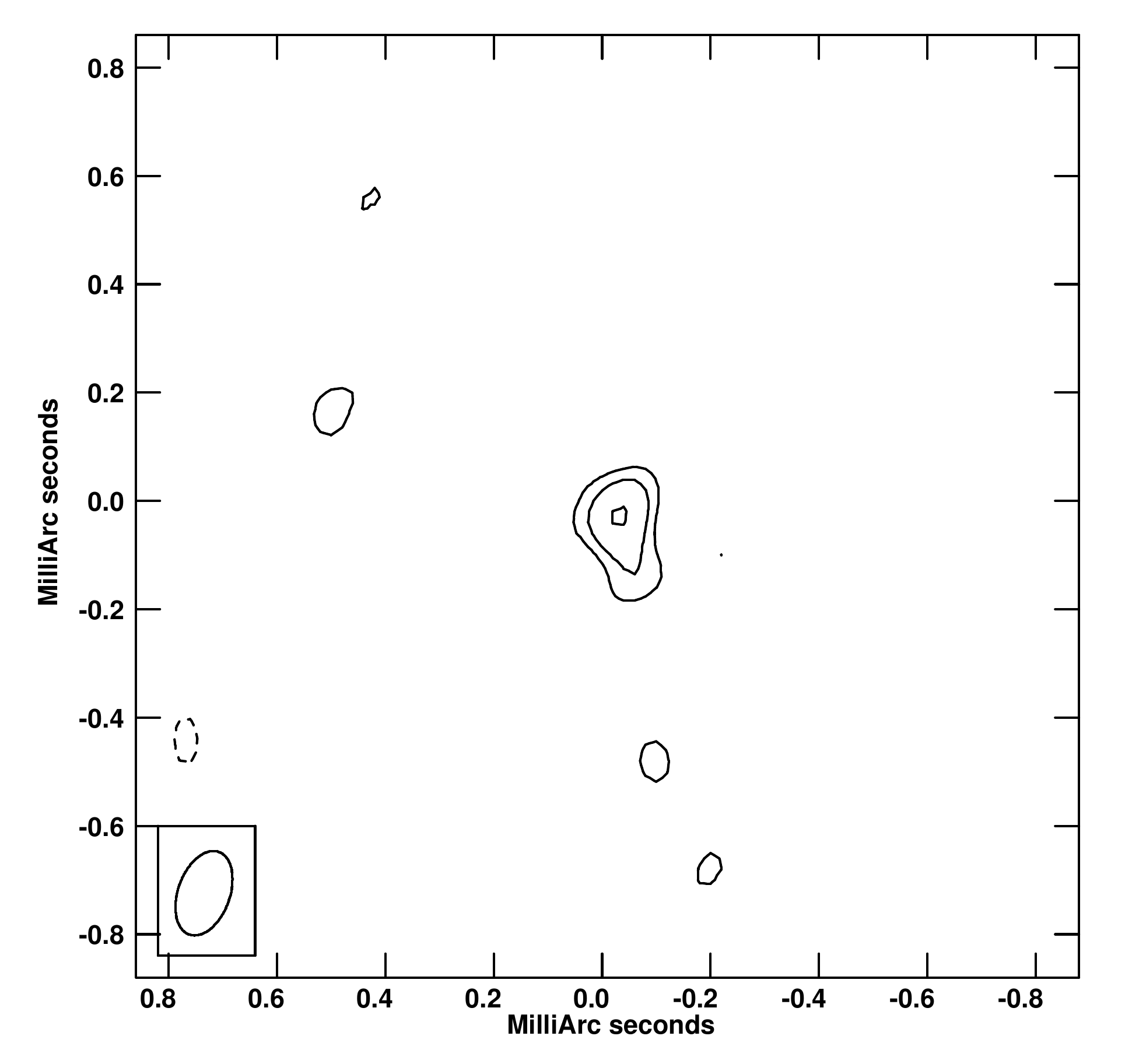}{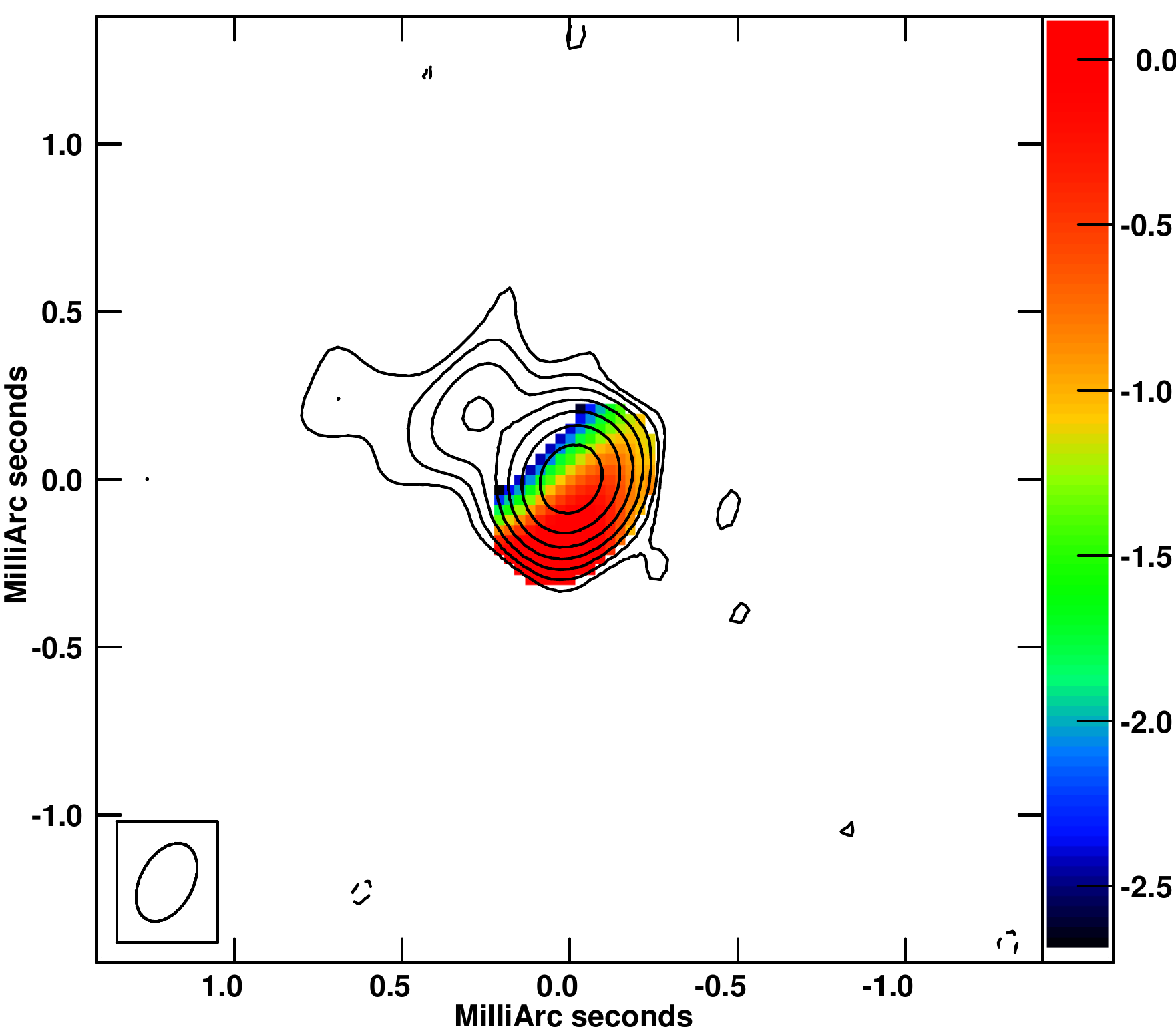}
\caption{Left, source-frequency phase referenced (SFPR-ed) image of M81* on 2016 Feburary 27 at 3.4\,mm. The peak intensity is 11.0 mJy\,beam$^{-1}$. The contour levels are 2.64$\times$($-$1, 1, 2, 4) mJy\,beam$^{-1}$. The lowest level is three times RMS of the image. The astrometric offset of the peak with respect to the phase center is  $-35\pm2$ $\mu$as in right ascension and $-32\pm3$ $\mu$as in declination, respectively. Right, spectral index map of M81* from 43.86 GHz (7\,mm) to 87.72 GHz (3.4\,mm) (grey scale flux range: $-$2.7 to 0.1), overlaid with 7\,mm total intensity contours (peak intensity 72.3 mJy\,beam$^{-1}$; lowest contour 0.69 mJy\,beam$^{-1}$). \label{fig:f4}}
\end{figure}



\end{document}